\documentstyle[twocolumn,epsf]{article}
\begin{document}

\renewcommand{\arraystretch}{1.0}
\renewcommand{\textfraction}{0.05}
\renewcommand{\topfraction}{0.9}
\renewcommand{\bottomfraction}{0.9}
\def\la{\;\raise0.3ex\hbox{$<$\kern-0.75em\raise-1.1ex\hbox{$\sim$}}\;}
\def\ga{\;\raise0.3ex\hbox{$>$\kern-0.75em\raise-1.1ex\hbox{$\sim$}}\;}
\def\lr{\;\raise0.3ex\hbox{$\rightarrow$\kern-1.0em\raise-1.1ex\hbox{$\leftarrow$}}\;}
\newcommand{\kB}{\mbox{$k_{\rm B}$}}           % k_B
\newcommand{\tn}{\mbox{$T_{{\rm c}n}$}}        % Tcn (crit.tem-re of neutrons)
\newcommand{\tp}{\mbox{$T_{{\rm c}p}$}}        % Tcp (crit.tem-re of protons)
\newcommand{\te}{\mbox{$T_{eff}$}}             % T effective
\newcommand{\dash}{\mbox{--}}                  % dash between numbers
\newcommand{\ex}{\mbox{\rm e}}                 % exponent (roman type)
\newcommand{\r}{\rule{0cm}{0.4cm}}
\newcommand{\hh}{\rule{0.5cm}{0cm}}
\newcommand{\hb}{\rule{0.4cm}{0cm}}
\newcommand{\rl}{\rule{0em}{1.8ex}}
\newcommand{\hhh}{\rule{1.4em}{0ex}}
\newcommand{\hhb}{\rule{1.2em}{0ex}}
\newcommand{\hhl}{\rule{2.4em}{0ex}}
\newcommand{\s}{$\;\;$}

\title{\bf A simple model of cooling neutron stars with
        superfluid cores: comparison with observations}
\author{K.P.~Levenfish, Yu.A.~Shibanov, D.G.~Yakovlev \\
        Ioffe Physical Technical Institute, Politekhnicheskaya 26 \\
        194021 St.-Petersburg, Russia 
        \footnote{e-mail: yak@astro.ioffe.rssi.ru}
        \footnote{To appear in Pisma v Astron.\ Zh.\ (Astron.\ Lett.)
        v.\ 25, No.\ 6, 1999 }
}
\date{}
\maketitle

\begin{abstract}
Cooling of neutron stars (NSs) with superfluid cores
is simulated taking into account neutrino emission
produced by Cooper pairing of nucleons.
Two cooling regimes of NSs
composed of matter with the same moderately
stiff equation of state are studied:
the standard cooling (for a $1.30 \, M_\odot$ star, as an example)
and the cooling enhanced by the direct Urca process
(for a $1.48 \, M_\odot$ star).
The critical temperatures of neutron and proton superfluidities,
\tn\ and $\tp$, are assumed to be constant over
the NS core, and treated as free parameters.
They are constrained using the surface temperatures
$T_s$ of isolated NSs
(RX J0822-43, PSR 1055-52, 1E 1207-52, Vela, Geminga,
PSR 0656+14, RX J0002+62),
obtained by interpretation of observed thermal radiation either
with black body spectrum or with hydrogen atmosphere models.

The temperatures $T_s$
given by the ``atmospheric" interpretation
for the last five objects
can be explained by cooling of the same NS,
with the same \tn\ and $\tp$ for all objects.
The confidence regions of \tn\ and $\tp$ are rather narrow
and depend on cooling regime.
Standard cooling requires moderately strong
neutron and proton superfluidities, while enhanced cooling
requires moderate neutron and strong proton superfluidities.
Adopting interpretation of observations with the black body
spectrum, we can propose similar explanation only for three
last objects. For the both,
standard and enhanced, cooling regimes
the confidence regions of $\tn$, $\tp$ appear to be wider
than in the case of 
the ``atmospheric'' values of $T_s$ and consist of two
separate subregions. None of the models requires
simultaneously strong neutron and proton superfluidities
($\tn,\, \tp \ga 3 \cdot 10^9$ K),
which is an argument against very soft equations of state in
the NS cores.
\end{abstract}

% Sec.1
\section{Introduction} % &&&&&&&&&&&&&&&&&&&&&&&&&&&&&&&&&&&&&&&&&&&&&&&&&

According to current theories, neutron stars (NSs) are born
hot, and then gradually cool. The first $10^5 $--$ 10^6$~yr
cooling proceeds via neutrino emission from the NS core, but later
it goes via photon surface emission.
Cooling depends on NS internal structure.
Thus, comparison of cooling theory with observational data
on NS surface temperatures enables one, in principle,
to explore NS structure, particularly, properties of
superdense matter in the NS cores. Among various properties
of this matter, the critical temperatures of
neutron and proton superfluidities,
$\tn$ and $\tp$, are of special interest. These quantities are
sensitive to parameters of strong nucleon--nucleon interaction
in superdense matter 
closely related to the main ``mystery" of NSs,
the equation of state in their cores (soft, moderate or stiff?).
The softer equation of state, the weaker nucleon--nucleon
repulsion at small separations, and the stronger
is nucleon superfluidity (i.e., higher $\tn$ and $\tp$).
The superfluidity affects the NS heat capacity and neutrino emission,
and thus the stellar cooling.
The aim of the present article is to continue our
studies (e.g., Levenfish \& Yakovlev 1996 and references therein)
of cooling of NSs with superfluid cores.
As in our previous articles, we assume, for simplicity, that the NS cores
consist of neutrons ($n$), with admixture of protons
($p$) and electrons ($e$).

Superfluidity was introduced into the cooling theory
by Tsuruta et al.\ (1972). However, it has not been
regarded as a powerful cooling regulator for a long time,
and it has been taken into account in a simplified manner
(as discussed, for instance, by Levenfish \& Yakovlev 1996).
Page and Applegate (1992) were the first who
emphasized the importance of superfluidity in
cooling of massive NSs, where
{\it enhanced} neutrino reactions of the direct Urca process
are allowed:
\begin{equation}
   n \to p + e + \bar{\nu}_e \, , \quad
   p+e \to n + \nu_e \, .
\label{Durca}
\end{equation}
The authors showed that the surface temperature of such a star
was almost solely determined by the values of \tn\ or \tp\
in its core (they assumed the superfluidity of one nucleon species)
in a wide interval of stellar ages,
$10^3$ -- $10^5$~yr. This means that
NSs are good ``thermometers" of nucleon superfluidity.
Moreover, the superfluidity appears to be a powerful regulator
of {\it standard} cooling
of less massive NSs (see, e.g., Page 1994, Levenfish \&
Yakovlev 1996)
determined by much weaker neutrino reactions of the modified
Urca-processes
\begin{equation}
  n + N  \to  p+N+e + \bar{\nu}_e \, , \;\;
  p + e + N  \to n + N + \nu_e \, ,
\label{Murca}
\end{equation}
and the nucleon-nucleon scattering
\begin{equation}
 N+N \to N+N + \nu + \bar{\nu}\, ,
\label{Brems}
\end{equation}
where $N$ is a nucleon ($n$ or $p$).

NS cooling is also greatly affected by
neutrino emission due to direct interband transitions of nucleons
whose dispersion relation contains a superfluid gap
(Schaab et al.\ 1997, Page 1998,
Yakovlev et al.\ 1998):
\begin{equation}
   N  \to N + \nu + \bar{\nu}.
\label{Cooper}
\end{equation}
This process can be treated as neutrino generation
due to {\it Cooper pairing of (quasi)nucleons}.
It has been included into the cooling theory quite recently.
Thus, previous cooling calculations have to be reconsidered.

Theoretical conclusion on the importance of superfluidity
in NS cooling was made at about the same time when
new observational data on NS surface
temperatures appeared. Below we present new cooling calculations
(Sect.\ 2) and compare them with observations
(Sect.\ 3).

% Sec. 2
\section{Cooling calculations}  %  &&&&&&&&&&&&&&&&&&&&&&&&&&&&&&&&&&&

% Sec. 2.1
\subsection{Cooling models} % ========================== *

Our cooling simulations have been described by Levenfish
\& Yakovlev (1996). Here we point out some details.
We use the models of NSs whose cores are composed
of matter with moderately stiff equation of state
proposed by Prakash et al.\ (1988)
(the version with the compression modulus $K_0=180$ MeV
and with the symmetry energy $S_V$ suggested by Page \&
Applegate 1992). The maximum NS mass, for a given equation of state, is
$1.73 \, M_\odot$. In order to study
the enhanced and standard
cooling we consider NS models with two masses $M$.
In the first case $M=1.48 \, M_\odot$, the NS radius is
$R=11.44$~km, and the central density
$\rho_c \! = \! 1.376 \cdot 10^{15}$~g/cm$^3$, while in the second
case $M=1.30 \, M_\odot$, $R=11.87$~km, and
$\rho_c \! = \! 1.07 \cdot 10^{15}$~g/cm$^3$. The adopted equation
of state allows the direct Urca process to operate at
densities
$\rho \! > \! \rho_{cr}\! = \! 1.30 \cdot  10^{15}$~g/cm$^3$.
%   rho_14=12.976
Thus the $1.48 \, M_\odot$ NS suffers the {\it enhanced}
cooling: the powerful direct Urca process (\ref{Durca})
is allowed in a small central kernel
of radius 2.32~km and mass $0.035 \, M_\odot$,
in addition to all the neutrino processes
(\ref{Murca})--(\ref{Cooper}) operating in the entire core.
On the other hand, the critical density $\rho_{cr}$
is not reached in the
$1.30\, M_\odot$ NS, and the star has the {\it standard}
neutrino luminosity determined by the processes
(\ref{Murca})--(\ref{Cooper}).
Notice that while composing the equation of state in our
previous cooling calculations the parameter
$n_0$ (saturation baryon number density)
has been set equal to 0.165 fm$^{-3}$.
In the present simulations, we set
$n_0=0.16$ fm$^{-3}$. That is why the NS mass
for the enhanced cooling models is slightly different
from that adopted earlier ($1.44\, M_\odot$).

Numerous microscopic calculations of Cooper pairing
in the NS cores (which use different models of
nucleon--nucleon interaction and 
manybody theories; see, e.g., Takatsuka \&
Tamagaki 1997 and references therein) predict a very large
scatter of superfluid critical temperatures of neutrons and
protons, \tn\ and \tp\ ($\sim 10^7$--$10^{10}$~K),
and their different density dependence. Therefore we
will make a simplified assumption that the critical temperatures
\tn\ and \tp\ are constant within the stellar core
and can be treated as free parameters. This is the main
simplification of the present article.
Let us assume that the proton superfluidity is produced by
the singlet-state
($^1{\rm S}_0$) pairing, while the neutron superfluidity
is due to the triplet-state
($^3{\rm P}_2$) pairing with zero projection
of total Cooper--pair momentum onto  quantization axis.
We will study cooling for different $T_{cn}$ and $T_{cp}$,
and determine those
$T_{cn}$ and $T_{cp}$ which are in better agreement with observations.

% Fig.1 %%%%%%%%%%%%%%%%%% FIGURE %%%%%%%%%%%%%%%%%%%%%%%%%%%%%%%%%%%%
\begin{figure}[t]                          % Fig.1
\begin{center}
\leavevmode
\epsfysize=8.5cm
\epsfbox{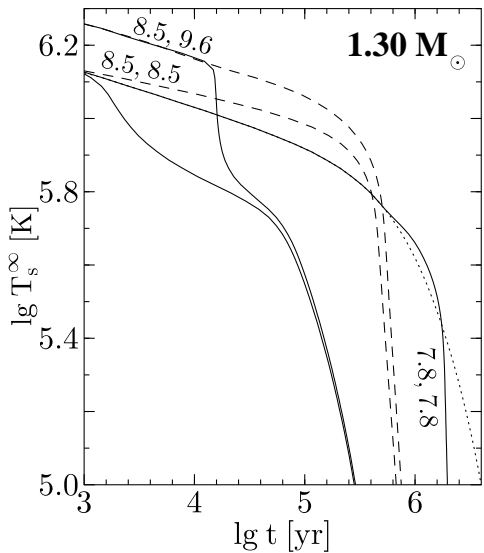}
\end{center}
\caption[]{
% Figure 1.
Effective surface temperature
$T_s^\infty$ (redshifted) versus age for standard cooling
of a NS ($1.30 \, M_\odot$) with the superfluid core.
Curves are labeled by the values of $\lg \tn$ and $\, \lg \tp$.
Solid and dashed lines are calculated including
and neglecting neutrino emission
produced by nucleon pairing, respectively.
The solid and dashed curves for
$\lg \tn=7.8$ and $\lg \tp=7.8$ coincide.
Dots show cooling of a nonsuperfluid star.\\[0.5ex]
}
\label{fig:rec_s}
\end{figure}

The heat capacity and neutrino luminosity in the reactions
(\ref{Durca})--(\ref{Brems})
in the superfluid
NS cores have been calculated as prescribed by
Levenfish \& Yakovlev (1996), while the luminosity in
the reaction (\ref{Cooper}) has been taken from
Yakovelv et al.\ (1998). We have also incorporated
the neutrino luminosity of the NS crust due to
bremsstrahlung emission by electrons which scatter off
atomic nuclei (using an approximate formula proposed
by Maxwell 1979). The NS heat capacity has been determined
as a sum of the capacities of $n$,
$p$ and $e$ in the stellar core; we have neglected
the heat capacity of the crust due to the smallness of the crust mass
in our NS models. While calculating the neutrino luminosity
and heat capacity the effective masses of neutrons and
protons in stellar matter, $m_n^\ast$ and $m_p^\ast$, have been
set equal to 0.7 of their bare masses.

%
% Fig.2 %%%%%%%%%%%%%%%%%% FIGURE %%%%%%%%%%%%%%%%%%%%%%%%%%%%%%%%%%%%
\begin{figure}[t]                          % Fig.2
\begin{center}
\leavevmode
\epsfysize=8.5cm
\epsfbox{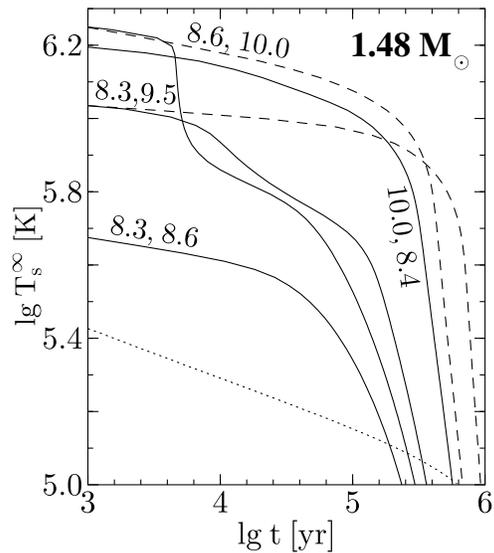}
\end{center}
\caption[]{
% Figure 2.
Enhanced cooling of a $1.48 \, M_\odot$ NS.
% Curves are labeled by the values of $\lg \tn$ and $ \, \lg \tp$.
% Solid and dashed lines are calculated
% including and neglecting
% Cooper-pairing neutrino emission, respectively.
Notations are the same as in Fig.\ 1.  % {fig:rec_s}
The solid and dashed lines for
$\lg \tn=8.3, \,\lg \tp=8.6$
and for $\lg \tn=10.0, \,\lg \tp=8.4$ coincide.
Dots show cooling of a nonsuperfluid star.
}
\label{fig:rec_e}
\end{figure}
%
%\newpage

As in our previous articles, we have used the NS cooling code
based on the approximation of isothermal interior. The effects
of General Relativity on NS structure and cooling have been
incorporated explicitly. The isothermal approximation
is justified for a NS of age
$t>(10$ -- $10^3)$~yr, when the internal thermal relaxation
is over, and only a narrow layer of the outer NS crust
is non-isothermal. In the isothermal region the quantity
$\;T_i(t)=T(r,t) \cdot \exp[\Phi(r)]$
(which can be treated as the internal temperature
corrected due to the local gravitational redshift)
is constant at any moment of time $t$.
Here, $T(r,t)$ is the real local temperature,
$\Phi (r)$ is the dimensionless gravitational potential, and
$r$ is a radial coordinate.
%At $\rho=\rho_b$ the local temperature
%is $T_b=T_i \, \exp(-\Phi_b)$
%(where $\Phi_b$ is the appropriate value of $\Phi$).

The photon NS luminosity depends on the effective
local surface temperature $T_s$.
The relationship between $T_s$ and the internal temperature
$T_i$ is determined by thermal insulation
of the outer stellar envelope. We have ignored
the effect of magnetic fields on NS cooling,
and have used the relationship
$T_s(T_i)$, obtained recently by Potekhin et al.\ (1997)
for $B=0$. We assume that the NS may possess a thin
surface layer
(of mass $\la 10^{-13} \, M_\odot$) of hydrogen or helium.
This layer cannot affect thermal insulation and
cooling but can change spectrum of thermal emission.
The effect of dipole surface magnetic fields
$B \la 5 \cdot 10^{12}$ G on NS cooling is
not large (Shibanov \& Yakovlev 1996). Therefore our
calculations can be applied, at least qualitatively,
for the NSs with such magnetic fields. In these cases,
$T_s$ means the mean effective surface temperature
which determines the total photon NS luminosity
$L_\gamma= 4 \pi \sigma R^2 T^{\,4}_s$
(non-redshifted),
where $\sigma$ is the Stefan--Boltzmann constant.
\vspace{-0.3cm}

% Sec 2.2 =======================================
\subsection{Effect of neutrino emission produced by
Cooper pairing of nucleons}

The main difference of the present simulations
from our previous ones is in including the neutrino
emission (\ref{Cooper}) due to Cooper pairing of nucleons.
The process was suggested by Flowers et al.\
(1976) and independently by Voskresensky \& Senatorov
(1987), but it has been included into the cooling
simulations only recently (see above).
The reaction rates have been recalculated and
improved by Yakovlev et al.\ (1998).
As shown by the latter authors, neutrino emission produced by
proton pairing is strongly suppressed due to
numerical smallness of weak vector neutral currents of
protons. The importance of the process for standard and
enhanced NS cooling is illustrated in Figs.\
\ref{fig:rec_s} and \ref{fig:rec_e}.

%
% Fig.3 %%%%%%%%%%%%%%%%%% FIGURE %%%%%%%%%%%%%%%%%%%%%%%%%%%%%%%%%%%%
\begin{figure}[t]                          % Fig.3
\begin{center}
\leavevmode
\epsfysize=8.5cm
\epsfbox{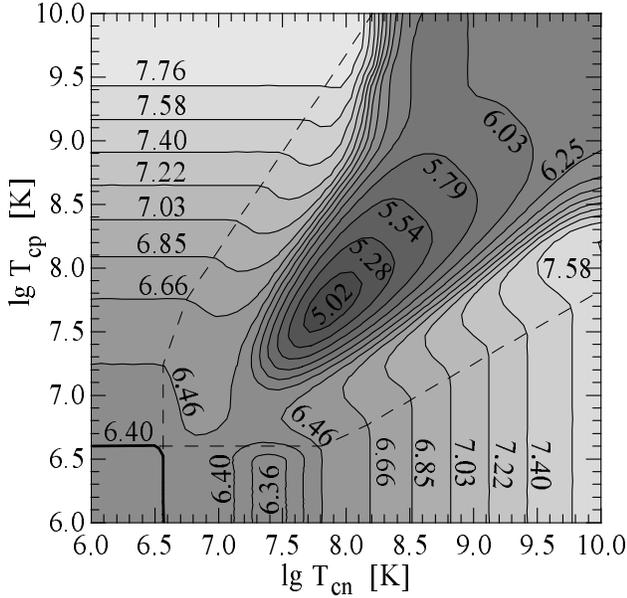}
\end{center}
\caption[]{
% Figure 3.
Lines of values of $T_{cn}$ and $T_{cp}$ which
correspond to certain internal temperatures
$T_i$ (the values of $\lg T_i$ are given near the curves)
or surface temperatures
$T_s^\infty$ (given in Table \protect{\ref{tab:ts_to_tc}})
of a NS with enhanced neutrino luminosity
($1.48 \, M_\odot$)
and Geminga's age ($3.4 \cdot 10^5$ yr). The darker the
background the cooler is the star. The region of joint
neutron and proton superfluidity (in the center and the upper
right corner) is enclosed by dashed lines.
A small region where the superfluidity has not
appeared at given age (the left lower corner) is separated
by the thick solid line.
}
\label{fig:Gem_2}
\end{figure}

It has been widely recognized that the superfluidity onset
at the neutrino cooling stage slows down the standard cooling.
As seen from Fig.\ \ref{fig:rec_s}, superfluidity,
on the contrary,
can accelerate the cooling. The acceleration
can be so large that by the end of the neutrino era
the surface temperature falls below the value it would
have for the enhanced cooling. This conclusion does not
apply to the cases of purely proton superfluidity
(proton pairing produces weak neutrino emission),
and strong purely neutron superfluidity
(neutrino emission due to neutron pairing dominates over
the standard neutrino processes
(\ref{Murca}) and (\ref{Brems}) only for
$T \la 10^9$~K, see Yakovlev et al.\ 1998).

%
% Fig. 4 %%%%%%%%%%%%%%%%%% FIGURE %%%%%%%%%%%%%%%%%%%%%%%%%%%%%%%%%%%%
\begin{figure}[t]
\begin{center}
\leavevmode
\epsfysize=8.5cm
\epsfbox{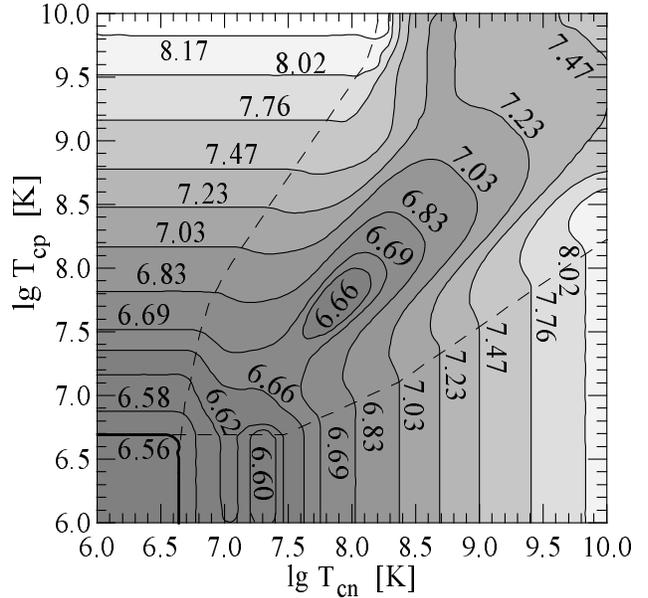}
\end{center}
\caption[]{
% Figure 4.
Same as in Fig.\ \protect{\ref{fig:Gem_2}}, but for a NS of the
PSR$\, 0656+14$ pulsar age ($10^5$ yr).
}
\label{fig:Gem_3}
\end{figure}

Neutrino emission due to pairing of neutrons
affects also enhanced cooling of NSs with superfluid
protons at $\tp \gg \tn$
(Fig.\ \ref{fig:rec_e}). In these cases, the proton superfluidity
appears in the NS much earlier than the neutron one, and it
has enough time to reduce the direct Urca process
by the beginning of the neutron pairing.
Therefore a splash of neutrino emission due to
the neutron pairing appears to be quite substantial;
the cooling does not slow down, as it would be
without Cooper neutrinos, but strongly speeds up.
\vspace{-0.3cm}

% Sec 2.3
\subsection{Results} % ====================================== *

We have calculated about 2000 cooling curves
which give the dependence of the effective NS surface temperature
$T_s^{\infty} = T_s \sqrt{1- R_g/R}$
(as detected by a distant observer, $R_g$ being the gravitational
radius) on age $t$.
For  $M=1.30 \, M_\odot$ and $M=1.48 \, M_\odot$ we have
$T^\infty_s/T_s= 0.822$ and 0.786, respectively.
The critical temperatures of neutrons \tn\
and protons \tp\ in the NS cores have been varied
in a wide interval from $10^6$ to $10^{10}$~K,
compatible with the scatter predicted by microscopic theories
(see Sect.\ 2.1).

%
% Fig. 5 %%%%%%%%%%%%%%%%%% FIGURE %%%%%%%%%%%%%%%%%%%%%%%%%%%%%%%%%%%%
\begin{figure}[t]
\begin{center}
\leavevmode
\epsfysize=8.5cm
\epsfbox{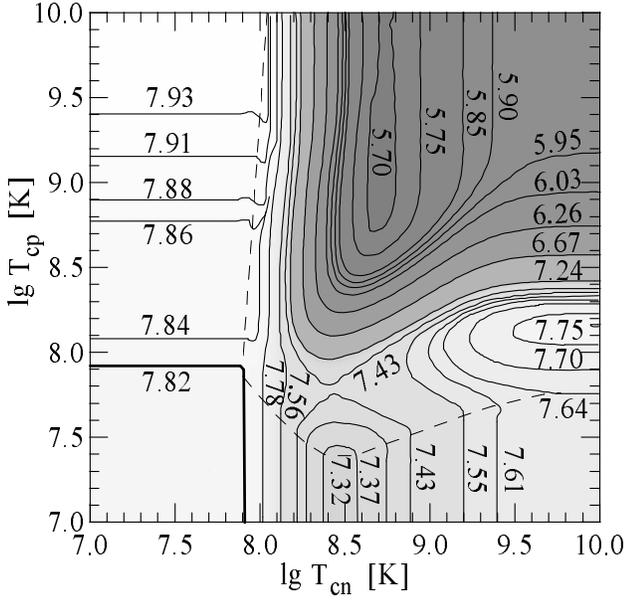}
\end{center}
\caption[]{
% Figure 5.
Isotherms of the internal temperature $T_i$ (or
the surface temperature $T_s^\infty$, Table
\ref{tab:ts_to_tc}) of a NS of the Geminga's age,
as in Fig.\ \protect{\ref{fig:Gem_2}}, but for
the standard cooling ($1.30 \, M_\odot$).
}
\label{fig:Gem_4}
\end{figure}
%

%
% Fig. 6 %%%%%%%%%%%%%%%%%% FIGURE %%%%%%%%%%%%%%%%%%%%%%%%%%%%%%%%%%%%
\begin{figure}[t]
\begin{center}
\leavevmode
\epsfysize=8.5cm
\epsfbox{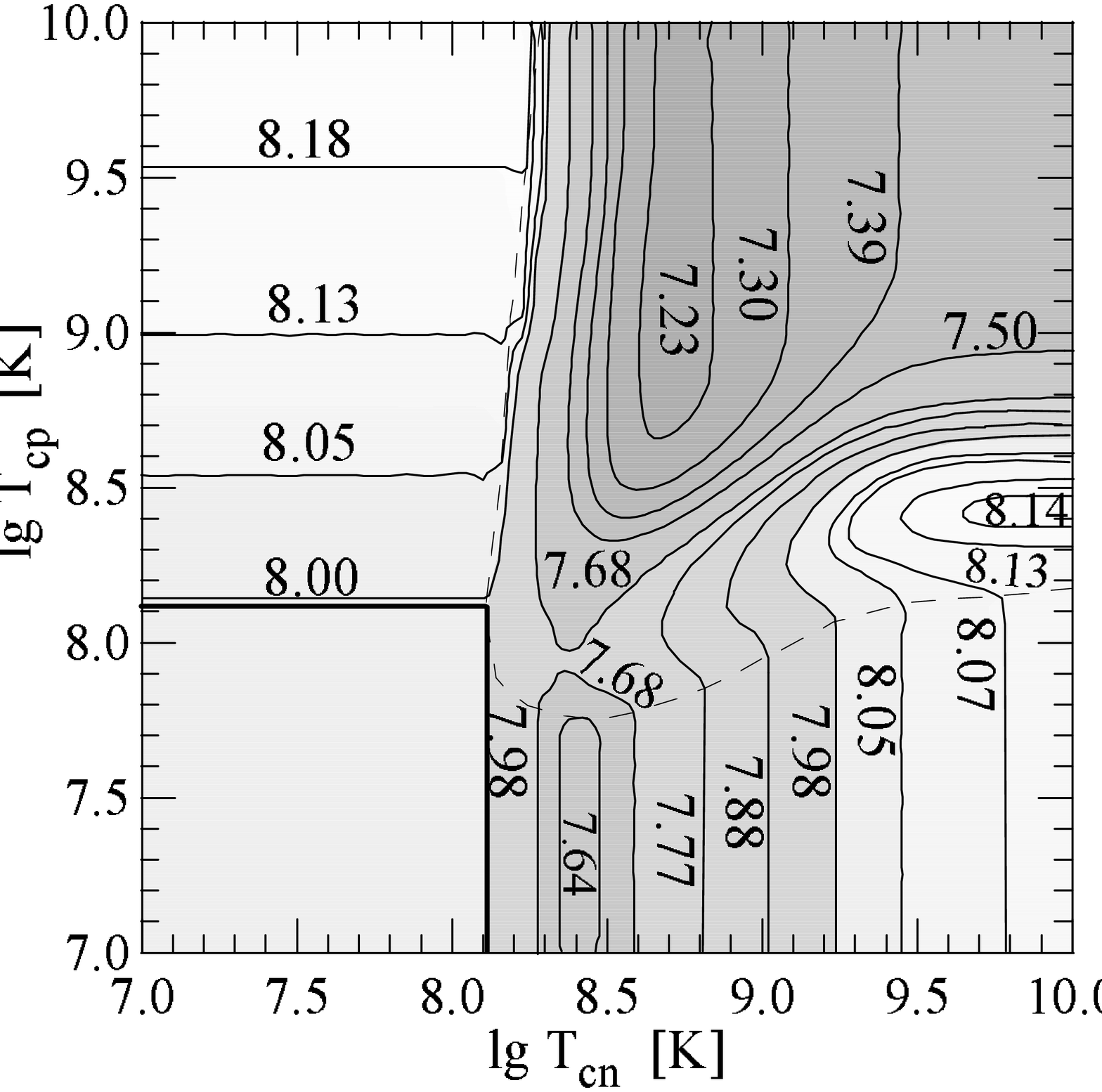}
\end{center}
\caption[]{
%Figure 6.
Same as in Fig.\ \protect{\ref{fig:Gem_4}}, but for a NS
of the PSR$\, 0656+14$ pulsar age.
}
\label{fig:Gem_5}
\end{figure}

Cooling curves for some selected
values of \tn\ and \tp\ are presented in Figs.\
\ref{fig:rec_s} and \ref{fig:rec_e}. However it is
inconvenient to analyse the results in the form of the cooling curves.
It is more convenient to plot those values of
($\tn,\; \tp $), which lead to a selected surface temperature
$T^\infty_s$ of the NS at given age $t$.
Figures \ref{fig:Gem_2}--\ref{fig:Gem_5} are arranged in this manner.
For example, we have chosen the values of $t$ which correspond
to the ages of the Geminga pulsar
($3.4 \cdot 10^5$~yr, Figs.\ \ref{fig:Gem_2} and \ref{fig:Gem_4})
and PSR$\,0656+14$ ($10^5$~yr, Figs.\ \ref{fig:Gem_3} and \ref{fig:Gem_5}).
Observational data on these and other cooling NSs are
given in Sect.~3.

In order to simplify understanding of the figures
let us outline the main properties of the heat capacity and
neutrino emission from superfluid NSs
(see Levenfish \& Yakovlev 1996, Yakovlev et al.\ 1998,
and references therein). The heat capacities of
$p$ and $n$ constitute, respectively,
$\sim 1/4$ and $ \sim 3/4$ of the total heat capacity $C_{tot}$.
Thus a strong proton superfluidity
reduces $C_{tot}$ by $\sim 25\%$, while a strong neutron superfluidity
reduces it by about 4 times.
If the ratio of the inner temperature
$T_i$ to the critical nucleon temperature
$T_{cN}$ in a cooling NS falls down from 0.3 to 0.1,
the heat capacity of nucleons
$N=n$ or $p$ decreases more than by three orders of magnitude,
and becomes much lower than the heat capacity of electrons.
Further reduction of the heat capacity of nucleons $N$
does not affect the total heat capacity.
The appearance of a weak superfluidity of nucleons $N$
almost doubles their heat capacity due to the
latent heat release at the phase transition.
The heat capacity of nucleons $N$ remains higher than
the heat capacity of normal nucleons when
$T_i/T_{cN}$ decreases from 1 to 0.3, i.e., as long as the difference
of temperature logarithms is $\lg T_{cN}
- \lg T_i \leq 0.5$.

The effects of superfluidities of $n$ and $p$ on the
neutrino reactions are also different.
For the reactions (\ref{Durca})--(\ref{Brems})
the difference is small. If we neglected neutrino
emission due to Cooper pairing of nucleons
the asymmetry of Figs.\ \ref{fig:Gem_2}--\ref{fig:Gem_5}
with respect to the inversion of axes $\tn \lr \tp$
would be mainly explained by different contributions of $n$ and
$p$ into the heat capacity. Cooper-pairing neutrinos
noticeably amplify the asymmetry: the neutrino emission
due to pairing of $p$ is weak, while the emission due to
pairing of $n$ dominates over the standard neutrino energy losses
at $T_i \la 10^9$ K $ \la \tn$ and over the direct Urca process
at $T_i \la \tn \ll \tp$. The main energy release in the process
occurs while
$T_i/T_{cN}$ decreases from 0.96 to $\sim 0.2$
(when the difference of temperature logarithms is
$\lg T_{cN} - \lg T_i \leq 0.7$).

Finally let us mention that for the enhanced cooling
the transition from the neutrino
to photon cooling stages
occurs somewhat later than for the standard cooling.
In a nonsuperfluid NS
which undergoes the standard cooling the neutrino era lasts
$t_\nu \sim $ (1--3) $\cdot 10^5$~yr.
A strong neutron superfluidity decreases
$t_\nu$ by a factor of 2--3, while a strong proton
superfluidity decreases $t_\nu$ by about
(20--30)$\%$.
If the both superfluidities, $n$ and $p$, are strong at once
$t_\nu \sim 10^4$ yr.
\\[0.2ex]

%\input{table1.tab}
% Relationship between the internal and surface temperatures
% for NS models with masses M=1.48Ms ¨ M=1.30Ms.

\begin{table*}[t]   % "*" ignores the twocoloumn-format if it's adopted
\caption[]{Relationships between $T_i$ and $T_s^{\infty}$ [K]
    for Figs.\ %3--6.
    \protect{\ref{fig:Gem_2}--\ref{fig:Gem_5}}.
    }
\label{tab:ts_to_tc}
\begin{center}
%====================================================================
\begin{tabular}{|p{0.8cm}|p{10.7cm}|}
\hline
\multicolumn{2}{|c|}{$ M= 1.48\, M_\odot^{\r}$}\\[1ex]
\hline
\parbox[t]{1.2cm}{lg$T_i^{\r}$\\[-0.3ex]
                  lg$T_s^\infty$\\[-1ex]
                  }
 &
\parbox[t]{12.7cm}{5.02\s 5.28\s 5.54\s 5.79\s 6.03\s 6.25\s 6.36\s 6.40\s 6.46\s 6.56\s 6.58\s 6.60\s 6.62  \\[-0.3ex]
                \s 4.50\s 4.60\s 4.70\s 4.80\s 4.90\s 5.00\s 5.05\s 5.07\s 5.10\s 5.15\s 5.16\s 5.17\s 5.18  \\[-1ex]
                  } \\
\hline
\parbox[t]{1.2cm}{lg$T_{i}^{\r}$\\[-0.3ex]
                  lg$T_s^\infty$\\[-1ex]
                  }
 &
\parbox[t]{15.8cm}{6.66 6.69\s 6.83\s 6.85\s 7.03\s 7.22\s 7.23\s 7.40\s 7.47\s 7.58\s 7.76\s 8.02\s 8.17  \\[-0.3ex]
                \s 5.20\s 5.22\s 5.29\s 5.30\s 5.40\s 5.50\s 5.51\s  5.60\s 5.64\s 5.70\s 5.80\s 5.95\s 6.03  \\[-1ex]
                  } \\
\hline
\end{tabular}\\[2ex]
% ===============================================================
\begin{tabular}[c]{|p{0.8cm}|p{9.8cm}|}
\hline
\multicolumn{2}{|c|}{$ M= 1.30\, M_\odot^{\r}$}\\[1ex]
\hline
\parbox[t]{0.8cm}{lg$T_i^{\r}$\\[-0.3ex]
                  lg$T_s^\infty$\\[-1ex]
                  }
 &
\parbox[t]{15.3cm}{5.70\s 5.75\s 5.85\s 5.90\s 6.95\s 6.03\s 6.26\s 6.67\s 7.23\s 7.24\s 7.30\s 7.32 \\[-0.3ex]
                \s 4.77\s 4.79\s 4.83\s 4.85\s 4.87\s 4.90\s 5.00\s 5.20\s 5.49\s 5.50\s 5.53\s 5.54 \\[-1ex]
 }\\
\hline
\parbox[t]{1.2cm}{lg$T_i^{\r}$\\[-0.3ex]
                  lg$T_s^\infty$\\[-1ex]
                  }
 &
\parbox[t]{15.8cm}{7.37\s 7.39\s 7.43\s 7.50\s 7.55\s 7.61\s 7.64\s 7.68\s 7.70\s 7.75\s 7.77\s 7.82\\[-0.3ex]
                \s 5.57\s 5.58\s 5.60\s 5.64\s 5.67\s 5.70\s 5.72\s 5.74\s 5.75\s 5.78\s 5.79\s 5.82 \\[-1ex]
 }\\
\hline
\parbox[t]{1.2cm}{lg$T_i^{\r}$\\[-0.3ex]
                  lg$T_s^\infty$\\[-1ex]
                  }
 &
\parbox[t]{15.8cm}{7.84\s 7.86\s 7.88\s 7.91\s 7.93\s 7.98\s 8.00\s 8.05\s 8.07\s 8.13\s 8.14\s 8.18 \\[-0.3ex]
                \s 5.83\s 5.84\s 5.85\s 5.87\s 5.88\s 5.91\s 5.92\s 5.95\s 5.96\s 5.99\s 6.00\s 6.02 \\[-1ex]
 } \\
\hline
\end{tabular}
\end{center}
\end{table*}

\noindent %##############################################################
{\bf Enhanced cooling} \\[0.2ex]
%
% Age     t2 = 1e5 yr   = 10^{5}
%         t1 = 3.4e5 yr = 10^{5.53}
%
Figures \ref{fig:Gem_2} and \ref{fig:Gem_3} illustrate
enhanced cooling of NSs of ages of the Geminga and
PSR$\,0656+14$ pulsars, respectively (see Table \ref{tab:NS_data} below).

Figure \ref{fig:Gem_2} shows isotherms of the
internal temperature $T_i$ of a NS of the Geminga's age
versus $\tn$ and $\tp$. Since
$T_i$ is strictly related to the surface temperature
these lines are also isotherms of
$T_s^\infty$ (Table \ref{tab:ts_to_tc}).
Dashes show auxiliary lines which enclose the region of
the joint superfluidity of nucleons.
The region to the left of the upper dashed line corresponds
to superfluidity of protons and normal neutrons, while
the region to the right of the lower line corresponds to
superfluidity of neutrons and normal protons.
The dashed lines intersect at the isotherm of temperature
$T_i=10^{6.4}$~K, to which a nonsuperfluid star would cool down
at moment $t$. This isotherm, plotted by the thick line,
encloses the region where the nucleon superfluidity
has not appeared by moment $t$ and does not affect the cooling.
Notice that owing to General Relativity effects 
isotherms $T_i$ correspond to somewhat higher local
temperatures of matter (see above). Thus, the isotherm
$T_i = 10^{6.4}$ K is associated with the values of
$T_{cN}$ which are slightly higher than $T_i$.

First of all we discuss the behaviour of
iso\-therms to the right of the lower auxiliary line.
The vertical parts of isotherms reveal that
NS cooling is governed by the only superfluidity of $n$.
With increasing $\tn$ at $\lg T_{cn} \ga 7.7$,
the core temperature $T_i$ grows up which can be explained like this.
The higher $\tn$, the earlier the neutron superfluidity appears;
accordingly the powerful direct Urca process is suppressed
earlier, cooling delay is longer, and the NS is
hotter at given $t$. For low $\tn$, the NS cools down
in a nonsuperfluid state the main part of its history, 
and is sufficiently cold at
age $t$. The neutrino luminosity of such a star is rather
weak and becomes comparable to the photon luminosity.
Additional neutrino energy losses due to 
neutron pairing at
$\lg \tn \ga 6.6$ slightly accelerate the cooling
at $6.6 \la \lg \tn \la 6.6 + 0.7$.
However at $\lg  \tn \la 6.6 + 0.5$ the acceleration
is compensated by the latent heat release produced by
the jump of the neutron heat capacity. The values
0.7 and 0.5 have been explained in the beginning of
this section.

Vertical parts of isotherms intersect
the auxiliary line at temperatures
$\tp$, at which the proton superfluidity appears.
With further increase of $\tp$,
a newly born proton superfluidity slightly decreases the
neutrino luminosity which is already reduced by
the strong neutron superfluidity.
The main effect will be caused by the heat capacity of
protons which jumps up. The latent heat release will increase
$T_i$: in a strip of width 
$\sim$0.5 (in $\lg \tp$) above the lower dashed line
the isotherms shift to the left. With further growth of
\tp\ the heat capacity is strongly reduced, the NS becomes
cooler and the isotherms shift to the right.

In a similar fashion, exchanging proton and neutron superfluidities,
we can explain horizontal parts of
the isotherms to the left of the upper auxiliary line.
In this case NS cooling in the range
$6.6 \la \lg \tp \la  6.6 +0.7$ does not appear because of
the weakness of the neutrino emission produced by
proton pairing. Strong proton superfluidity suppresses
the heat capacity weaker than neutron superfluidity
with the same critical temperature. Therefore,
at high \tp\ the star is warmer than for the same
\tn\ at vertical parts of isotherms
in the lower right part of the figure.

The horizontal parts of the isotherms intersect
the upper auxiliary line at temperatures
$\tn$ which switch on the neutron superfluidity.
This superfluidity induces the latent heat release
and associated slight NS heating (dips on isotherms
to the right of the auxiliary line). With increasing
\tp\ the NS heating produced by neutron superfluidity 
decreases and disappears (the dips vanish).
The effect is caused by the neutrino emission
due to neutron pairing which is more pronounced
if the direct Urca process is strongly suppressed.
At high $\tp$ the neutron pairing becomes the main neutrino
emission mechanism. Since the neutrino emission due to
neutron pairing is more efficient than
the latent heat release the appearance of the neutron superfluidity
with the growth of \tn\ does not delay the cooling but, on the contrary,
accelerates it.
A noticeable NS cooling
via Cooper-pairing neutrinos, takes place
in a strip of width 0.7 (in $\lg \tn$) to the right of
the upper dashed line. Lower and to the right of this strip
the cooling is associated with the reduction of
the neutron heat capacity.

Now consider the region of
$\tn$, $\tp$ between the auxiliary lines in more detail.
Some increase of $T_i$ at
$\lg \tn \la 6.6 +0.5$ and $\lg \tp \la 6.6 + 0.5$
is caused by the latent heat release at onset
of weak superfluidity of $n$ and $p$. Further growth of $\tn  \sim  \tp$
induces initially rapid decrease and then weak increase
of $T_i$. The decrease is explained by exponential reduction
of the heat capacity by the joint superfluidity of $n$ and $p$
while the weak increase is associated with reduction of the
direct Urca process at early cooling stages.

Figure \ref{fig:Gem_3} is analogous to Fig.\ \ref{fig:Gem_2},
but corresponds to a younger NS of age $t=10^5$~yr.
Comparing Figs.~\ref{fig:Gem_2} and \ref{fig:Gem_3} we see
that one should have $\tp \ll \tn$ or $\tn \ll \tp$
to support high surface temperature $T_s^\infty$ (or $T_i$)
for a longer time.\\[0.2ex]

\noindent %#######################################################
{\bf Standard cooling}\\[0.2ex]
%
% Age     t2 = 1e5 yr   = 10^{5}
%         t1 = 3.4e5 yr = 10^{5.53}
%
Standard cooling of a $1.30\, M_\odot$ NS is illustrated in
Figs.~\ref{fig:Gem_4} and \ref{fig:Gem_5} for the stars of
Geminga and PSR $\, 0656+14$ ages, respectively.
Isotherms are qualitatively different from
those for enhanced cooling
(cf.\ with Figs.\ \ref{fig:Gem_2} and \ref{fig:Gem_3}):
even an approximate symmetry of the neutron and proton
superfluidities is absent. The asymmetry is attributed to
the weakness of the standard neutrino energy losses.
First, in the absence of such a powerful cooling regulator as
the direct Urca process the difference
of the heat capacities of $n$ and $p$ (see above) is more pronounced.
Second, Cooper-pairing neutrino emission becomes
more important on the background of the weaker neutrino emission
produced by other neutrino reactions; Cooper-pairing emission 
is asymmetric itself being more efficient for neutrons 
than for protons.

If the superfluidity is absent, a NS
($1.30 \, M_\odot$) enters the photon cooling stage
at $t_\nu \sim 1.6 \cdot 10^5$~yr.
Thus the PSR$\, 0656+14$ pulsar appears at the transition stage
and Geminga is at the photon stage.
The neutrino luminosity is already weak and 
superfluidity affects the cooling mainly either through
the heat capacity or through the Cooper-pairing neutrino emission.

Consider, for instance,  Fig.\ \ref{fig:Gem_4}. 
Two dashed auxiliary lines enclose the domain of joint superfluidity of nucleons.
To the left of the upper line protons are superfluid and
neutrons not, while below the lower line only neutrons are
superfluid. The lines intersect at the isotherm of the temperature
$ \lg T_i \approx 7.82$, which a nonsuperfluid NS 
would have at age $t$.
The superfluidity with
$\lg \tn \la 7.9$ and $\lg \tp \la 7.9$ has not appeared
by moment $t$ and does not affect the cooling.

Horizontal parts of isotherms to the left of the upper dashed line
show that cooling is regulated by the proton superfluidity alone.
At $ \lg \tp \ga 7.9$~K the superfluidity appears just before
given moment $t$ and is weak. In the range $7.9 \la \lg \tp \la 7.9 + 0.5$
it initiates the latent heat release and a weak increase of
$T_i$. At higher  $\lg \tp \ga 8.4$ the proton heat capacity is
reduced, and the total NS heat capacity decreases by $\sim 25 \%$.
On the other hand, for high  $\tp$, the proton superfluidity onset
is shifted to the neutrino cooling stage. The cooling delay produced by
suppression of the neutrino luminosity in the neutrino era
is somewhat stronger than the cooling
acceleration produced by the effect of superfluidity on the
heat capacity. Thus $T_i$ continues its growth with
increasing $\tp$.

% Fig. 7 %%%%%%%%%%%%%%%%%% FIGURE %%%%%%%%%%%%%%%%%%%%%%%%%%%%%%%%%%%%
\begin{figure}[t]
\begin{center}
\leavevmode
\epsfysize=8.5cm
\epsfbox{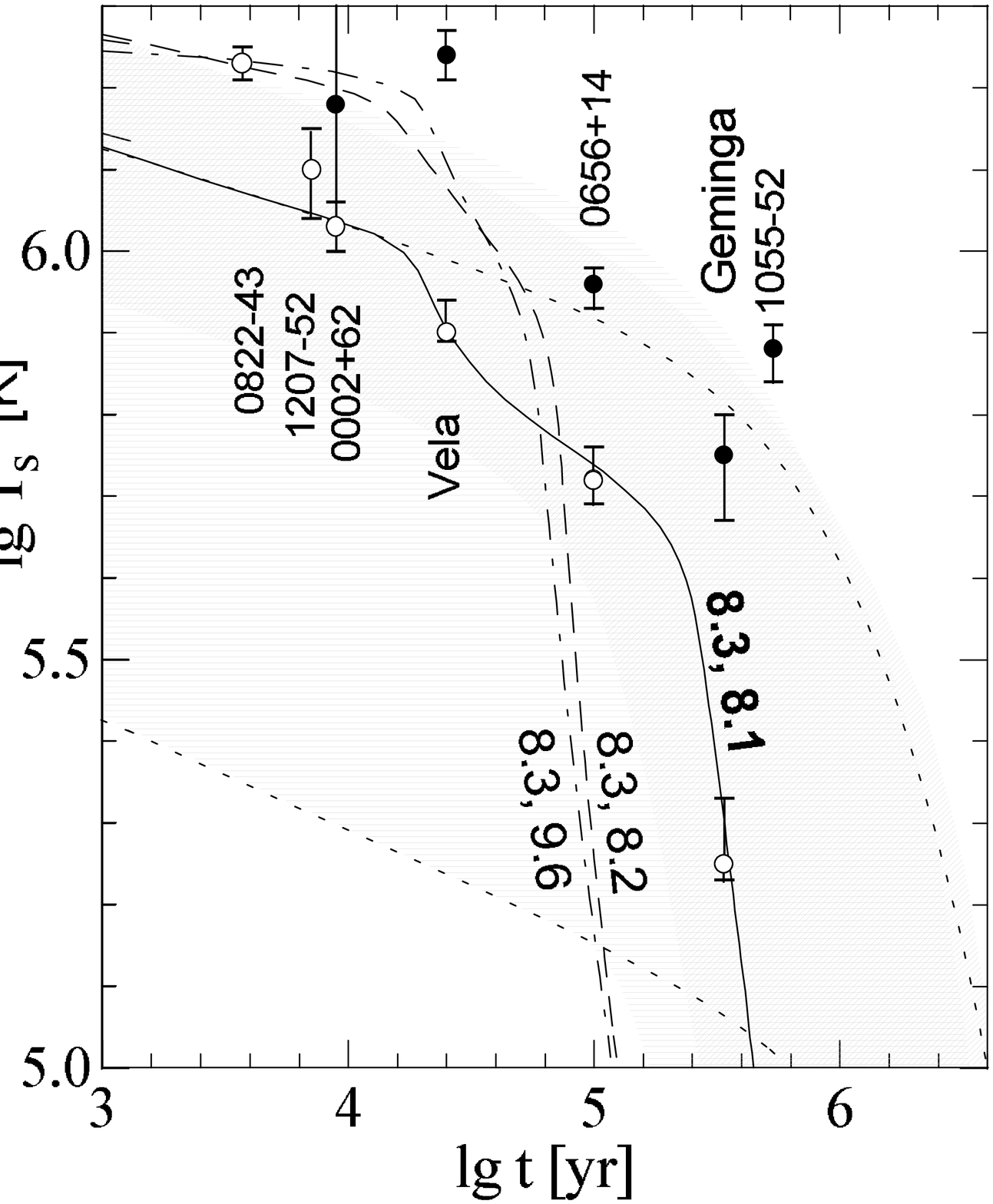}
\end{center}
\caption[]{
% Figure 7.
Observational data on surface temperatures of NSs obtained
(Table 2) in the models of
black body radiation (filled circles) and hydrogen atmosphere
(open circles). Shaded regions show ranges of $T_s^\infty$
filled by standard (diagonal shading)
and enhanced (horizontal shading) cooling curves
of NSs with different \tn\ and $\tp$ (from $10^6$ to $10^{10}$~K).
The solid line shows standard cooling ($1.30 \, M_\odot$)
at specified values $\lg \tn$ and $\, \lg \tp$ (given near the curve).
Dotted lines exhibit standard (upper line)
and enhanced (lower line) cooling of a non-superfluid NS.
The dashed and dot-and-dashed lines show, respectively,
standard and enhanced cooling of a NS which possesses
an envelope of mass
$7\cdot 10^{-10} M_\odot$ composed of light elements
(see Sect.\ 3.3).
}
\label{fig:NS_data}
\end{figure}

Horizontal parts of isotherms end at temperatures
$\tn \sim 10^8$~K, at which the neutron superfluidity is switched on.
The neutron pairing induces a splash of neutrino emission and 
cooling acceleration in a strip of the width 
$\sim$0.7 (in $\lg \tn$) to the right of the auxiliary line.
The minimum of $T_i$ takes place in the interval
$\lg \tn \approx 8.5$--$8.7$,
because at $8.0 \la \lg \tn \la 8.0 + 0.5 $ the neutrino energy losses
are partly compensated by the latent heat release. The lowest temperatures
$T_i$ are realized at $\tp \gg 10^8$~K,
when the nucleon heat capacity suffers the strongest suppression.

\begin{table*}[t]%[pht]   % "*" ignores the twocoloumn-format if it's adopted
\caption{Observational data }
\label{tab:NS_data}
\begin{center}
\begin{tabular}{|p{2.2cm}|p{1.5cm}|@{}p{12.3cm}|}
\hline                
\hline
% Column titles ===============================
 Source
    &
 $\;\;\lg t\rl$ [yr]
    & \begin{tabular}[c]{@{$\;\;\;$}p{3.2cm}|p{1.9cm}|p{2.3cm}|p{3.5cm}@{}}
 Model$^{\,b)}$
    &
 $\;\;\lg T_{\!s}^{\infty}\;$ [K]\rl
    &
  Confidence level$^{\,a)}$
    &
 Reference \rule{0em}{2.5ex}
                  \end{tabular} \\
%==================================================
\hline
\hline
RX$\,$J0822-43 & \hhh 3.57 &
  \begin{tabular}[c]{@{$\;\;\;$}p{3.2cm}|p{1.9cm}|p{2.3cm}|p{3.5cm}@{}}
   Hydrogen atm.\      &\hhb $6.23^{+0.02\rl}_{-0.02} $
                       &\hhl 95.5\%
                       & Zavlin et al.\ (1999b) \\[0.5ex]
  \hline
            Black body & $\hhb 6.61^{+0.05\rl}_{-0.05}  $
                       &\hhl 95.5\%
                       & Zavlin et al.\ (1999b) \\[0.5ex]
  \end{tabular}    \\
  \hline
        1E$\,$1207-52  &\hhh 3.85 &
  \begin{tabular}[c]{@{$\;\;\;$}p{3.2cm}|p{1.9cm}|p{2.3cm}|p{3.5cm}@{}}
   Hydrogen atm.\      &\hhb $6.10^{+0.05\rl}_{-0.06} $
                       &\hhl $\,$90\%
                       & Zavlin et al.\ (1998) \\[0.5ex]
  \hline
            Black body &\hhb $6.49^{+0.02\rl}_{-0.01}  $
                       &\hhl $\,$90\%
                       & Zavlin et al.\ (1998) \\[0.5ex]
  \end{tabular}    \\
  \hline
        RX$\,$J0002+62 & \hhh $3.95^{c)}$ &
  \begin{tabular}[c]{@{$\;\;\;$}p{3.2cm}|p{1.9cm}|p{2.3cm}|p{3.5cm}@{}}
   Hydrogen atm.\      &\hhb  $6.03^{+0.03\rl}_{-0.03} $
                       &\hhl 95.5\%
                       & Zavlin et al.\ (1999a) \\[0.5ex]
  \hline
   Black body &\hhb $6.18^{+0.18\rl}_{-0.18} $
                       &\hhl 95.5\%
                       & Zavlin et al.\ (1999a) \\[0.5ex]
  \end{tabular}    \\
  \hline
  \begin{tabular}{@{}l}
        PSR~0833-45 \\
        (Vela)
  \end{tabular}
                       & \hhh $4.4^{d)}$ &
  \begin{tabular}[c]{@{$\;\;\;$}p{3.2cm}|p{1.9cm}|p{2.3cm}|p{3.5cm}@{}}
   Hydrogen atm.       &\hhb $5.90^{+0.04\rl}_{-0.01}$
                       &\hhl $\,$90\%
                       & Page et al.\ (1996) \\[0.5ex]
  \hline
   Black body &\hhb $6.24^{+0.03\rl}_{-0.03} $
                       &\hhl $\;\;\;$---
                       & \"{O}gelman (1995)\\[0.5ex]
  \end{tabular}    \\
  \hline
   PSR~0656+14 & \hhh 5.00 &
  \begin{tabular}[c]{@{$\;\;\;$}p{3.2cm}|p{1.9cm}|p{2.3cm}|p{3.5cm}@{}}
   Hydrogen atm.       &\hhb $5.72^{+0.04\rl}_{-0.02} $
                       &\hhl $\;\;\;$---
                       & Anderson et al.\ (1993) \\[0.5ex]
  \hline
   Black body &\hhb $5.96^{+0.02\rl}_{-0.03} $
                       &\hhl $\,$90\%
                       & Possenti et al.\ (1996) \\[0.5ex]
  \end{tabular}    \\
  \hline
  \begin{tabular}{@{}l}
     PSR~0630+178\\
     (Geminga)
  \end{tabular}
                       & \hhh 5.53 &
  \begin{tabular}[c]{@{$\;\;\;$}p{3.2cm}|p{1.9cm}|p{2.3cm}|p{3.5cm}@{}}
   Hydrogen atm.       &\hhb  $5.25^{+0.08\rl}_{-0.01} $
                       &\hhl $\,$90\%
                       & Meyer et al.\ (1994) \\[0.5ex]
  \hline
   Black body &\hhb $5.75^{+0.05\rl}_{-0.08} $
                       &\hhl $\,$90\%
                       & Halpern, Wang (1997) \\[0.5ex]
  \end{tabular}    \\
  \hline
  PSR~1055-52 & \hhh 5.73 &
  \begin{tabular}[c]{@{$\;\;\;$}p{3.2cm}|p{1.9cm}|p{2.3cm}|p{3.5cm}@{}}
  Black body &\hhb $5.88^{+0.03\rl}_{-0.04} $
                       &\hhl $\;\;\;$---
                       & \"{O}gelman (1995) \\[0.5ex]
  \end{tabular} \\
  \hline
  \multicolumn{3}{@{}l@{}}{
  \begin{tabular}{@{}l@{}}
  \rule{0em}{3ex}$^{a)}\,\rl${\footnotesize
        Confidence level of $T_s^\infty$ (90\% and 95.5\% correspond
        to $1.64\sigma$ and $2\sigma$, respectively);
        dash means that the level is not indicated}\\[-0.7ex]
  $\phantom{^{ a) }}\,\rl${\footnotesize
        in cited references.
        }\\[-0.7ex]
    $^{b)}\,$\rl{\footnotesize
         Model used for interpretation of observation.
         }\\[-0.7ex]
  $^{c)}\,$\rl{\footnotesize
       The mean age taken according to Craig et al.\ (1997).
       }\\[-0.7ex]
  $^{d)}\,$\rl{\footnotesize According to Lyne et al.\ (1996). }
\end{tabular}
 }\\
\end{tabular}
\end{center}
\end{table*}

On vertical parts of isotherms below the lower dashed line
the cooling is regulated by the neutron superfluidity alone.
With growing \tn\ in this domain the temperature
$T_i$ varies in a different manner than with growing
\tp\ in the domain of the purely proton superfluidity.
This happens because at $ \lg \tn \ga 7.9$ the neutrino emission
due to the neutron pairing is important.
It speeds up NS cooling in the range
$7.9 \la \lg \tn \la 7.9 + 0.7 $ (see above).
If $7.9 \la \lg \tn \la$ 7.9 + 0.5 the neutrino cooling is
partly compensated by the latent heat release. Therefore
the minimum of $T_i$ takes place in the interval
$7.9+0.5 \la \lg \tn \la 7.9+0.7$.
It is not so deep as in the upper part of Fig.\
\ref{fig:Gem_4} since the nucleon heat capacity is suppressed only partly.

A strong neutron superfluidity reduces the heat capacity stronger
and the neutrino luminosity weaker than a strong proton superfluidity.
Owing to the weakness of the neutrino energy losses
this difference is sufficient for a NS with high \tn\ and
normal $p$ to cool in a different way than at equally
high \tp\ and normal $n$. Strong neutron superfluidity
delays the cooling of those NSs which would be at the neutrino cooling stage
or at the neutrino-photon transition stage in case they were nonsuperfluid.
This is demonstrated in Fig.\ \ref{fig:Gem_5} 
(for PSR$\, 0656+14$);
in the absence of superfluidity, at $t=10^5$~yr this pulsar
would be at the transition stage. Strong neutron superfluidity
accelerates cooling of older NSs, e.g.,
of the Geminga's age (Fig.\ \ref{fig:Gem_4}).

Vertical parts of isotherms in Fig.\
\ref{fig:Gem_4} intersect the auxiliary line at temperatures
$\tp$, at which the proton superfluidity appears.
This superfluidity leads to the latent heat release
and to the growth of $T_i$ in a strip of width
$\sim$0.5 in $\lg \tp$ above the lower dashed line.
With further growth of \tp\ the heat capacity is strongly
reduced and cooling is accelerated. For a very strong joint
superfluidity of $n$ and $p$
($ \tn \gg 10^{7.9}$~K, $\tp \gg 10^{7.9}$~K, the very right upper corner
of Fig.~\ref{fig:Gem_4}), the nucleon heat capacity and
the neutrino luminosity of the NS core are fully suppressed,
and the cooling is governed by the electron heat capacity.

Figure \ref{fig:Gem_5} is analogous to Fig.~\ref{fig:Gem_4},
but corresponds to a younger NS. Its neutrino luminosity
is somewhat higher and the relative contribution
of neutron--pairing neutrinos is smaller.
Neutrino emission produced by pairing
affects the cooling weaker.

% Sec. 3 &&&&&&&&&&&&&&&&&&&&&&&&&&&&&&&&&&&&&&&&&&&&&&&&&&&&&&&&&&&&&&&&&
\section{Comparison of calculations and observations}

% Sec. 3.1 =================================================
\subsection{Observations and their interpretation}
By now, thermal X-ray radiation has been detected from seven
isolated NSs (Table \ref{tab:NS_data}).
All the sources are reliably identified.
Four of them (Vela, Geminga, PSR~0656+14 and PSR~1055-52)
are radiopulsars. The other three NSs
(1E~1207-52, RX~J0002+62 and RX J0822-43) are radio silent,
but are observed as point-like sources of soft X-rays in
supernova remnants. Their identification as isolated NSs
is confirmed by observations of coherent pulsations
of X-ray emission with NS-spin periods
(excluding 1E~1207-52) and smooth
(modulation depth $\la 15-40\%$)
light curves, and also by small ratios of the optical to
X-ray luminosities, which are not typical for accreting NSs in binaries.

%
% Fig. 8 %%%%%%%%%%%%%%%%%% FIGURE %%%%%%%%%%%%%%%%%%%%%%%%%%%%%%%%%%%%
\begin{figure*}[t]
\begin{center}
\leavevmode
\epsfysize=8.5cm
\epsfbox{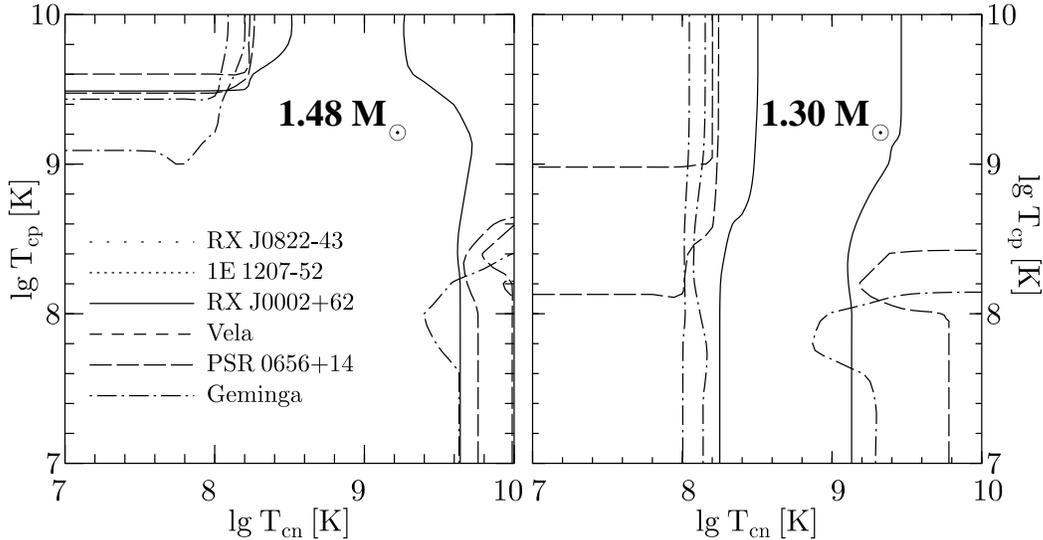}
\end{center}
\caption[]{
% Figure 8.
Confidence regions of $T_{cn}$ and $T_{cp}$ which
correspond to the ``black body" surface temperatures
(Table 2,  Fig.\ \protect{\ref{fig:NS_data}})
of RX J0002+62, PSR~0656+14 and Geminga
in the models of
enhanced (left) and standard (right) NS cooling.
}
\label{fig:all_bb}
\end{figure*}

As a rule, an X-ray spectrum from an isolated NS contains two components.
The spectrum of the softer component is similar to the black body spectrum
for a temperature $T_s \approx
10^5$--$10^6$~K. According to the cooling theory, 
the surface of a middle-aged NS, $t \approx 10^3 $--$ 10^6$ yr,
should have about the same temperature. It is assumed, therefore,
that the main contribution into the soft component
comes from the thermal surface emission.
The hard component is associated either with a thermal
emission from hot polar caps on the pulsar surface
(heated by an inverse current of energetic particles from the
magnitosphere) or with a non-thermal magnetospheric emission.

The thermal NS radiation is often fitted by the black body
spectrum. However, the interpretations based on the spectra,
which are obtained by modelling of the NS atmospheres, are more realistic.
At the same effective temperature
$T_s$ the black body spectrum appears to be softer.
Therefore, while interpreting the same observations,
the ``black body'' temperature
$T_s$ is noticeably
(typically, by a factor about 1.5--3) higher than the ``atmospheric''
temperature.

% Fig. %%%%%%%%%%%%%%%%%% FIGURE %%%%%%%%%%%%%%%%%%%%%%%%%%%%%%%%%%%%
\begin{figure*}[t]
\begin{center}
\leavevmode
\epsfysize=8.5cm
\epsfbox{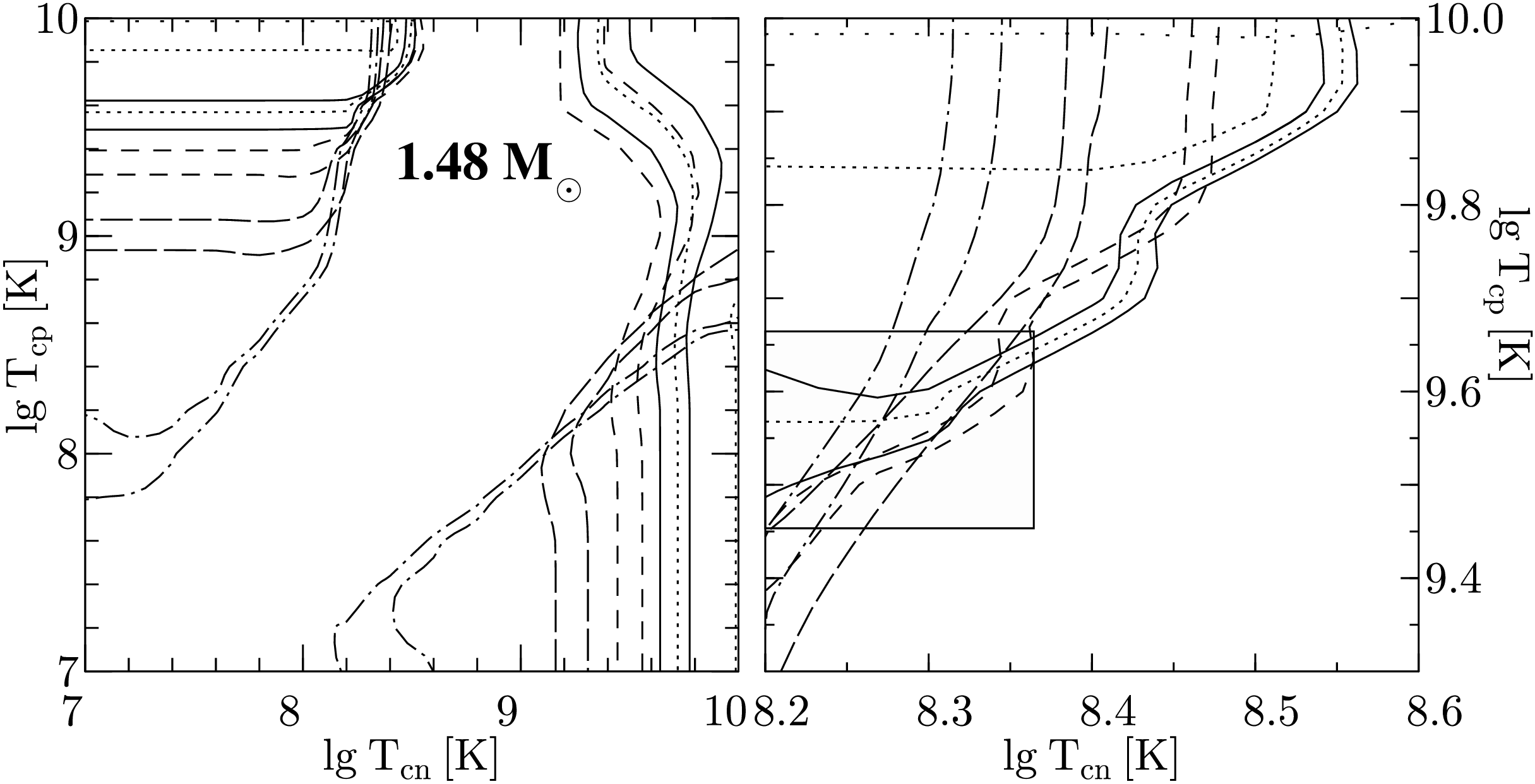}
\end{center}
\caption[]{
% Figure 9.
Same as in Fig.\ \protect{\ref{fig:all_bb}}, but for the enhanced
cooling and ``atmospheric" interpretation
(Table 2) of spectra of thermal
radiation from RX~J0822-43, 1E~1207-52, RX~J0002+62, Vela,
PSR~0656+14 and Geminga. On the right panel
we show in more detail the region (shaded rectangle)
in which the allowed values of $T_{cn}$ and $T_{cp}$
of five latter objects  are either close or intersect.
}
\label{fig:fast_atm}
\end{figure*}

Construction of the atmosphere models in a practically important
range of temperatures and surface magnetic fields $B$ has not
been completed yet. Only the hydrogen-helium and iron
atmosphere models with weak magnetic fields $B \la 10^9$ G
are really reliable as well as the hydrogen-helium models
with strong fields ($B \ga 10^{11}$ G) for
$T_s \ga 10^6$ K. If these conditions are fulfilled, the thermal
radiation spectrum and the NS light curves are better
described by the atmospheric model than by the black body model.
Physical parameters inferred from atmospheric interpretations
(radiating surface area, distance to a NS, column density of interstellar
gas, etc.) are in better agreement with the data
obtained from independent observations in other wavelengths
(see, e.g., Page et al.\ 1996,
Zavlin et al.\ 1998, 1999a,b). The black body interpretation
often gives less realistic values of the parameters and
meets some difficulties in explaining all sets of observational
data.

In Table \ref{tab:NS_data} the NSs, the sources of the thermal emission,
are ordered according to their ages. For every NS, we present
its age, the method of interpretation of thermal radiation,
and the effective surface temperatures
$T_s^\infty=T_s \sqrt{1-R_g/R \,}$
obtained by a given method.  For the pulsars, we present
the dynamic ages, while for the radio silent objects we give
the ages determined from the associated supernova remnants.
Although the values of $M$ and $R$ (and, therefore, of the gravitational
redshift) used in our models
(Sect.\ 2) differ slightly from those obtained
(or adopted) from interpretation of observations,
the temperatures $T_s^\infty$ from Table \ref{tab:NS_data}
can be compared with our cooling calculations due to
the following reasons. First, our calculations are not very
sensitive to variations of NS mass or radius
(see Sect.\ 3.3 below). Second, the confidence ranges
of $M$ and/or $R$, obtained (adopted) from interpretation of
observations, include, as a rule, our theoretical values.

Among the atmospheric interpretations presented in Table
\ref{tab:NS_data}, the results for 1E~1207-52,
RX~J0002+62, RX~J0822-43 and for the Vela pulsar seem to
be more reliable. The advantage of interpretation of their spectra
by the hydrogen-helium atmosphere models is additionally confirmed
by the fact that fitting of observations by the
iron atmosphere models or by the black body spectrum is
of lower statistical significance.
The Geminga and PSR~0656+14 pulsars are older and cooler.
The important contribution into the spectral opacity of their
atmospheres should come from the effects of motion of neutral
and partly ionized atoms across magnetic fields
(Potekhin \& Pavlov 1997, Bezchastnov et al.\ 1998).
However, only approximate atmosphere models have been used
for interpretation so far, in which these effects
have been ignored.
The presence of the hydrogen atmosphere of the Geminga pulsar
is additionally confirmed by a possible discovery
of the proton cyclotron line in the optical spectrum
(Bignami et al.\ 1996, Martin et al.\ 1998).
The ``atmospheric'' temperatures of the Geminga and PSR 0656+14 pulsars
are likely to be closer to the reality than the black body temperatures,
but less reliable than for younger NSs.

We will assume that the atmospheric and black body interpretations
are equally possible and analyze the values of
\tn\ and \tp\ which can be inferred from these data.

% Sec. 3.2 ============================================================== *
\subsection{Necessity of superfluidity for interpretation
           of observations}

In Fig.\ \ref{fig:NS_data} the cooling curves are compared with
the data of Table \ref{tab:NS_data}. Diagonal and horizontal
shadings show the ranges of the surface temperature
$T_s^\infty(t)$ filled, respectively, by
different standard and enhanced cooling
curves calculated by varying
\tn\ and \tp\ from $10^6$ to $10^{10}$ K.
Dashes show cooling of nonsuperfluid stars.

The ``nonsuperfluid'' curves are seen to be in poor
agreement with observations. On the other hand, the observations
can be explained by assuming superfluidity in the NS cores.
This is illustrated (Fig.\ \ref{fig:NS_data}) by the
standard cooling curve (the solid line).
The values of \tn\ and \tp\ are chosen in such a way
to hit the maximum number of observational points at once.

According to Fig.\ \ref{fig:NS_data},
all the ``atmospheric'' temperatures as well as the
``black body'' temperatures of RX~J0002+62, PSR~0656+14 and Geminga
hit the allowed regions of the standard and enhanced cooling of
superfluid NSs. Thus our cooling calculations can be
compared with the ``atmospheric" and ``black body" temperatures
$T_s$ of these sources.

High black body surface temperatures of
RX~J0822-43, 1E~1207-52 (not presented in
Fig.\ \ref{fig:NS_data}, but given in Table
\ref{fig:NS_data}), Vela, and PSR~1055-52 are not explained by
our models but can be explained by other models of cooling NSs,
including the models with superfluid cores. For instance,
observations of the Vela pulsar agree with the standard
cooling of a superfluid NS
($\tn=10^7$~K,$\,\tp=10^{10}$~K) possessing an outer envelope
of mass $\sim 10^{-9}\, M_\odot$ composed of light elements
(Potekhin et al.\ 1997). High black body temperatures of
RX~J0822-43, 1E~1207-52 and PSR~1055-52 may indicate the
presence of some additional reheating mechanism inside these
sources (see, e.g., Umeda et al.\ 1993) or the presence of
superstrong ($B \ga 10^{14} $~G) magnetic fields (Heyl \& Hernquist 1997).
Finally, one cannot exclude the possibility that the black body
interpretation of their spectra is incorrect (see above).

% Sec. 3.3
\subsection{Results} % ====================

As seen from Fig.\ \ref{fig:NS_data}, by assuming the presence
of nucleon superfluidity in the NS cores, we can explain
the majority of NS observations either by 
standard or by enhanced NS cooling.
These observations include six ``atmospheric" interpretations
(RX~J0822-43, 1E~1207-52, RX~J0002+62, Vela, PSR~0656+14, Geminga)
and three ``black body'' ones (RX J0002+62,
PSR~0656+14, Geminga). For simplicity, we assume that
the internal structure of these NSs is the same, and, particularly,
the NSs have the same mass. Then the critical temperatures
of nucleons in their cores are the same.
Let us analyse the confidence regions of
\tn\ and $\tp$ constrained by observations.
Including observations of several NSs allows us to reduce
these regions.

%
% Fig. %%%%%%%%%%%%%%%%%% FIGURE %%%%%%%%%%%%%%%%%%%%%%%%%%%%%%%%%%%%
\begin{figure*}[t]
\begin{center}
\leavevmode
\epsfysize=8.5cm
\epsfbox{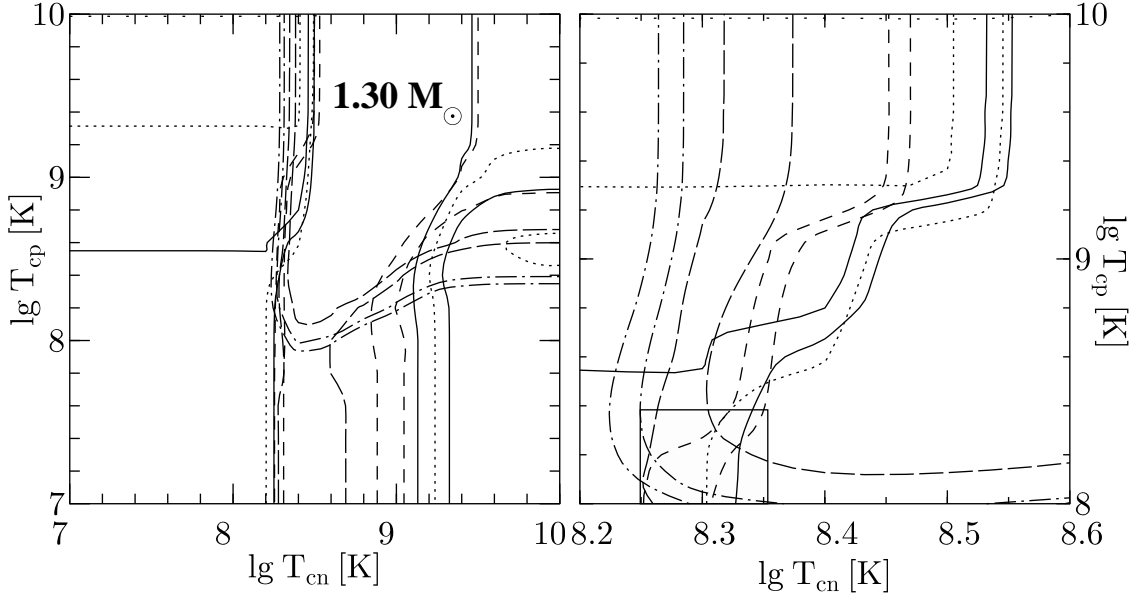}
\end{center}
\caption[]{
% Figure 10.
Same as in Fig.\ \protect{\ref{fig:fast_atm}},
but for standard cooling.
}
\label{fig:std_atm}
\end{figure*}

The regions in question are plotted in Figs.\
\ref{fig:all_bb}--\ref{fig:std_atm}.
Figure \ref{fig:all_bb} corresponds to the standard and
enhanced cooling of NSs
(with masses $1.30\, M_\odot$ and $1.48 \, M_\odot$, respectively)
with the black body spectrum.  Figure
\ref{fig:fast_atm} corresponds to the enhanced cooling
of the stars
($1.48 \, M_\odot$) possessing hydrogen atmospheres.
Finally, the standard cooling of NSs ($1.30 M_\odot$) with
hydrogen atmospheres is shown in Fig.\ \ref{fig:std_atm}.
In every figure, the lines of different type
enclose the confidence regions of $T_{cn}$ and $T_{cp}$ 
associated with the error bars of the
observed NS surface temperatures
$T_s^\infty$ (Table 2). Correspondence of the
lines to the selected NSs is displayed in Fig.\ \ref{fig:all_bb}.
For the  PSR~0656+14 and Geminga pulsars, isotherms are taken
from Figs.\ \ref{fig:Gem_2}--\ref{fig:Gem_5}. In each figure
the actual overall confidence region of \tn\ and \tp\ lies at the overlap
of the confidence regions of all objects.

Figures \ref{fig:all_bb}--\ref{fig:std_atm} show that
observations of several NSs at once can be explained
either by the standard or by the enhanced cooling,
adopting either black body or atmospheric interpretations of the
spectra. In all the cases there are the
ranges of \tn\ and $\tp$ close or joint for all NSs; they do not
contradict to the microscopic theories of nucleon
superfluidity in NSs (see Takatsuka \& Tamagaki 1997
and references therein).

According to the left panel of Fig.\ \ref{fig:all_bb}, 
by adopting the enhanced cooling and the
black body interpretation of observations we obtain two
confidence regions of \tn\ and $\tp$; 
each explains cooling of three
objects at once. The first region corresponds to a moderate neutron superfluidity
($\lg \tn \approx 8.1$) and a strong proton superfluidity
($\lg \tp \approx 9.5$); the second, wider region
corresponds to a strong neutron superfluidity
($\lg \tn \approx  9.75$--10.0) and a moderately weak proton superfluidity
($\lg \tp < 8.3$). For the standard cooling of a NS with the
black body spectrum (the right panel of Fig.\ \ref{fig:all_bb})
there are also two regions of $\tn$ and $\tp$ for the three NSs,
but they are somewhat different than for the enhanced cooling.
The first region corresponds to a moderate neutron superfluidity
($\lg \tn = 8.0$--8.1) and a moderately strong
proton superfluidity ($\lg \tp = 8.2$--$9.0$), while the
second one is associated with a strong
neutron superfluidity 
($\lg \tn > 9.8$) and a moderately weak proton one
($\lg \tp < 8.2$).

In the case of the enhanced cooling and the ``atmospheric"
surface temperatures (Fig.\ \ref{fig:fast_atm})
there is the only region of joint or very close values of
\tn\ and \tp\ for five NSs
(the shaded rectangle on the right panel). It corresponds
to a moderate neutron superfluidity
($\lg \tn \approx 8.2$--8.35) and a strong proton superfluidity
($\lg \tp \approx 9.45$--9.65). This superfluidity
($\lg \tp =8.3,\,\lg \tp = 9.6$) allows us to describe the sixth
object from the ``atmospheric'' set (RX~J0822-43),
if we assume that it possesses a thermally insulating envelope
of light elements with the mass equal to $ 7\cdot 10^{-10} M_\odot$
(see Fig.\ \ref{fig:NS_data}). However even in the absence of
the envelope the confidence region of
\tn\ and \tp\ for this object is sufficiently close
to the joint confidence region indicated above.

Finally, for the standard cooling and the ``atmospheric" spectrum
(Fig.\ \ref{fig:std_atm}), there is again the only 
confidence region of \tn\ and $\tp$,
where the critical temperatures are
nearly the same for the five NSs;
it corresponds to a moderately strong superfluidity of $n$ and $p$
($\lg \tn \approx 8.25$--8.35, $\lg \tp \approx 8.0$--8.4,
shaded rectangle). The observations of
RX~J0822-43 can also be explained by the presence of the
same superfluidity
($\lg \tn = 8.3$, $\lg \tp = 8.2$) assuming that the object
possesses an envelope of light elements of the 
$7\cdot 10^{-10} M_\odot$ mass (see Fig.\ \ref{fig:NS_data}).
In the absence of the envelope, the confidence region of
$\tn$ and $\tp$ lies outside 
of the joint confidence region 
since the error bar of $T_s^\infty$ for this object
is situated at the boundary of the shaded region (filled by 
the cooling curves of superfluid NSs).

Therefore, the constraints on the critical temperatures
\tn\ and $\tp$ depend on cooling type and on interpretation
of the NS thermal radiation. The fact that the NSs have
actually different masses has little effect on our results
as long as we assume that
\tn\ and \tp\ are independent of density.
The standard cooling curves of NSs with masses lower
than the threshold mass $M_{cr}=1.442\, M_\odot$
(of switching on the direct Urca process) practically coincide.
The same is true for the enhanced cooling curves of NSs
($M \ga M_{cr}$) as long as their cores contain a moderate
or strong superfluidity of
$n$ and/or $p$ with $T_c \ga 10^8$~K. With weakening of the
superfluidity the difference between enhanced and standard
cooling curves increases reaching maximum for nonsuperfluid NSs.

The existence of the same critical temperatures
for several NSs at once is quite unexpected.
We have expected different joint confidence regions of
$T_{cn}$ and $T_{cp}$ for different pairs of NSs.
The result is even more surprising taken into account
simplicity of our cooling models. It would be interesting
to confirm (or reject) these results using advanced
cooling models and a larger number of objects.
 
As a whole, one needs stronger superfluidity of at least
one nucleon species for the enhanced cooling models.
%This indicates indirectly that at densities
%$\rho \approx (4$--$5) \, \rho_0$, at which the direct
%Urca process becomes allowed, the critical temperature
%of nucleons of given species should be higher
%than for lower $\rho$. 
In the case of the black body
interpretation of NS radiation, the stellar cores may contain
a strong neutron superfluidity,
$\tn \approx 10^{9.7}$--$10^{10.0}$~K. In neither case
one needs strong superfluidity of $n$ and $p$ simultaneously
to explain the observations. This indicates that the equation
of state of the NS cores cannot be too soft
(the softness would mean weak nucleon--nucleon repulsion at small separations
which would induce especially strong pairing).

Finally let us mention that we can satisfy
observations by varying the only parameter, $\tp$,
if we assume the presence of neutron superfluidity
with $\tn \approx 10^{8.1}$--$10^{8.3}$~K. For the standard
cooling, this parameter should lie in the moderate range
$10^{8.0}$--$10^{9.0}$~K, while for the enhanced cooling
it should be larger, $10^{9.45}$--$10^{9.65}$~K.

% Sec. 4
\section{Conclusions} % &&&&&&&&&&&&&&&&&&&&&&&&&&&&&&

Our results show that cooling of NSs is very sensitive
to superfluid state of neutrons and protons in their cores.
This confirms the possibility (Page \& Applegate 1992)
to explore the nucleon superfluidity in the NS cores, the fundamental
property of matter of supranuclear density, from NS cooling.
Our calculations confirm also that the neutrino emission due to
nucleon pairing plays important role in NS cooling.

We have used the simplified NS models with two fixed masses,
$1.48 \, M_\odot$ and $1.30 \, M_\odot$, and nucleon critical
temperatures constant over the stellar cores.
Even in this case we have been able to explain the majority
of observations of the thermal radiation from
Vela, Geminga, PSR~0656+14, RX J0002+62, 1E~1207-52 by four models:
by the standard or enhanced cooling of NSs using the black body or
``atmospheric" interpretations of their spectra. For every model,
we have obtained its own, sufficiently constrained 
confidence regions of the critical temperatures
$T_{cn}$ and $T_{cp}$.

Our results are preliminary. For obtaining more reliable conclusions
on superfluidity and equation of state in the NS cores,
one needs new cooling simulations (for different model equations
of state and nucleon critical temperatures
taking into account variation of
$T_{cn}$ and $T_{cp}$ over the NS cores),
new high-quality observations of NS thermal emission and
new atmospheric models for interpretation of observations.

We are grateful to G.G.~Pavlov and V.E.~Zavliv
for providing us with the data on
RX~J0822-43 and RX~J0002+62 prior to publication.
The work is partly supported by the grants
of RFBR (96-02-16870a), DFG-RFBR (96-02-00177G) and INTAS (96-0542). \\[3ex]

\noindent
{\bf\Large References}
\begin{itemize}
\renewcommand{\labelitemi}{}
\itemindent=-0.7cm
\itemsep=0cm
\item %bibitem{acprt93}
   Anderson~S.B., C\'{o}rdova~F.A.,
    Pavlov~G.G., Robinson~C.R., Thompson~R.J.
    ApJ, 1993, 414, 867.

\item %\bibitem{bpv98}
   Bezchastnov~V.G., Pavlolv~G.G., Ventura~J.
   Phys.\ Rev.\ 1998, A58, 180.

\item %\bibitem{bcmeb96}
   Bignami~G.F., Caraveo~P.A., Mignani~R., Edelstein~J., Bowyer~S.
   ApJ (Lett.) 1996, 456, L111.

\item %\bibitem{chp97}
   Craig~W.W., Hailey~Ch.J., Pisarski~R.L.
   ApJ 1997, 488, 307.

\item %\bibitem{frs76}
   Flowers~E.G., Ruderman~M., Sutherland~P.G.
   ApJ 1976, 205, 541.

\item %\bibitem{hw97}
   Halpern~J., Wang~F.  ApJ 1997, 477, 905.

\item %\bibitem{hh97}
  Heyl~J., Hernquist~L.
  ApJ (Lett.) 1997, 489, L67.

\item %\bibitem{lpgc96}
   Lyne~A.G., Pritchard~R.S., Graham-Smith~F., Camilo~F.
   Nature 1996, 381, 497.

\item %\bibitem{ly96}
   Levenfish~K.P., Yakovlev~D.G.
   Astron.\ Lett.\ 1996, 22, 47.

% \item %\bibitem{lsy98}
%   Levenfish~K.P., Shibanov~Yu.A., Yakovlev~D.G.
%   Physica Scripta 1998, T77, 79.

\item %\bibitem{maxwell79}
   Maxwell~O.V. ApJ 1979, 231, 201.

\item %\bibitem{mhs98}
   Martin~C., Halpern~J.P., Schiminovich~D.
   ApJ (Lett.) 1998, 494, L211.

\item %\bibitem{mpm94}
   Meyer~R.D., Pavlov~G.G., M\'{e}sz\`{a}ros~P.
   ApJ 1994, 433, 265.

\item %\bibitem{ogelman95}
   \"{O}gelman~H. In: Lives of Neutron Stars
   (eds.\ Alpar~M.A.\ et al.), Dordrecht:
   Kluwer 1995, p.\ 101.

\item %\bibitem{page94}
   Page~D. ApJ 1994, 428, 250.

\item %\bibitem{page98}
   Page~D. In: The Many Faces of Neutron Stars
   (eds.\ Buccheri~R., van Paradijs~J., Alpar~M.A.).
   Dordrecht: Kluwer 1998, p.\ 538.

\item %\bibitem{pa92}
   Page~D., Applegate~J.H.
   ApJ (Lett.) 1992, 394, L17.

\item %\bibitem{psz96}
   Page~D., Shibanov~Yu.A.,  Zavlin~V.E. In:
   R\"{o}ntgenstrahlung from the Universe (MPE Report No.263)
   (eds.\ Zimmermann~H.U., Tr\"{u}mper~J.E., Yorke~H.).
   Garching: Max-Planck Institute f\"{u}r
   Extraterrestrische Physik 1996, p.\ 173.

\item %\bibitem{pmc96}
   Possenti~A., Mereghetti~S., Colpi~M.
   A\&A 1996, 313, 565.

\item %\bibitem{pcy97}
   Potekhin~A.Y., Chabrier~G., Yakovlev~D.G.
   A\&A 1997, 323, 415.

\item %\bibitem{pp97}
   Potekhin~A.Y., Pavlov~G.G.
   ApJ 1997, 483, 414.

\item %\bibitem{pal88}
   Prakash~M., Ainsworth~T.L., Lattimer~J.M.
   Phys.\ Rev.\ Lett.\ 1988, 61, 2518.

\item %\bibitem{tt97}
   Takatsuka~T., Tamagaki~R.
   Progr.\ Theor.\ Phys.\ 1997, 97, 345.

\item %\bibitem{usnt93}
   Umeda~H., Shibazaki~N.,  Nomoto~K., Tsuruta~S.
   ApJ 1993, 408, 186.

\item %\bibitem{tclr72}
  Tsuruta~S., Canuto~V., Lodenquai~J., Ruderman~M.
  ApJ 1972, 176, 739.

\item %\bibitem{svsww97}
   Shaab~C., Voskresensky~D., Sedrakian~A.D., Weber~F., Weigel~M.K.
   A\&A 1997, 321, 591.

\item %\bibitem{sy96}
   Shibanov~Yu.A., Yakovlev~D.G. A\&A 1996, 309, 171.

\item %\bibitem{vs87}
   Voskresensky D., Senatorov A.
   Sov.\ J.\ Nucl.\ Phys. 1987, 45, 411.

\item %\bibitem{ykl98}
   Yakovlev~D.G., Kaminker~A.D., Levenfish~K.P.
   In: Neutron Stars and Pulsars (ed.\ Shibazaki~N. et al.).
   Tokyo: Universal Academy Press 1998, p.\ 195.

\item %\bibitem{zpt98}
   Zavlin~V.E., Pavlov~G.G., Tr\"{u}mper~J.
   A\&A 1998, 331, 821.

\item %\bibitem{zpt99}
   Zavlin~V.E., Pavlov~G.G., Tr\"{u}mper~J.
%  RX J0002+6242
   ApJ 1999a, in press.

\item %\bibitem{ztp99}
   Zavlin~V.E., Tr\"{u}mper~J., Pavlov~G.G.
%  RX 0822-43
   1999b (in preparation, to be submitted to ApJ).

\end{itemize}
 
\end{document}